\documentclass{article}

\usepackage{arxiv}

\usepackage[utf8]{inputenc} 
\usepackage[T1]{fontenc}    
\usepackage{hyperref}       
\usepackage{url}            
\usepackage{booktabs}       
\usepackage{amsfonts}       
\usepackage{nicefrac}       
\usepackage{microtype}      

\usepackage{graphicx}
\usepackage{doi}

\usepackage{comment}
\usepackage[flushleft]{threeparttable}
\usepackage{array}
\usepackage{multirow}
\usepackage{tikz}
\usetikzlibrary{decorations.pathreplacing}
\usetikzlibrary{spy}
\usepackage{colortbl, booktabs}
\usepackage{amsmath,amssymb,amsfonts}
\graphicspath{{./Images/}}
\usepackage{textcomp}
\usepackage{xcolor}
\usepackage{algorithmic}
\definecolor{Gray}{gray}{0.9}
\newcolumntype{P}[1]{>{\centering\arraybackslash}p{#1}}
\def\BibTeX{{\rm B\kern-.05em{\sc i\kern-.025em b}\kern-.08em
    T\kern-.1667em\lower.7ex\hbox{E}\kern-.125emX}}

\title{Community Detection in Cryptocurrencies with Potential Applications to Portfolio Diversification}

\author{ {\includegraphics[scale=0.06]{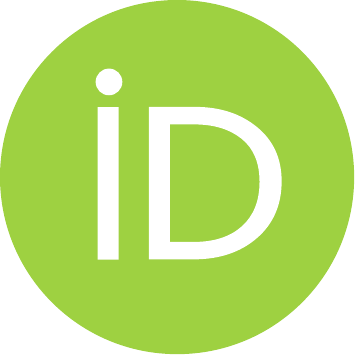}\hspace{1mm}Jenna Gavin} \\
	School of Computing\\
	Dublin City University\\
	Glasnevin, Dublin  9\\
	\texttt{jenna.gavin6@mail.dcu.ie} \\
	\And
	\href{https://orcid.org/0000-0001-7598-3126}{\includegraphics[scale=0.06]{orcid.pdf}\hspace{1mm}Martin Crane} \\
	School of Computing\\
	Dublin City University\\
	Glasnevin, Dublin  9\\
	\texttt{martin.crane@dcu.ie} \\
}

\renewcommand{\shorttitle}{Community Detection in Cryptocurrencies}

\hypersetup{
pdftitle={Community Detection in Cryptocurrencies},
pdfsubject={q-fin.CP, q-fin.ST},
pdfauthor={Jenna Gavin, Martin Crane},
pdfkeywords={Financial Markets, Graph Theory, Cryptocurrencies, Correlation Matrix, Community Structures, Principal Component Analysis},
}

\begin{document}
\maketitle

\begin{abstract}
In this paper, the cross-correlations of cryptocurrency returns are analysed. The paper examines one years’ worth of data for $146$ cryptocurrencies from the period 01/01/2019 to 31/12/2019. The cross-correlations of these returns are firstly analysed by comparing eigenvalues and eigenvector components of the cross-correlation matrix C with Random Matrix Theory (RMT) assumptions. Results show that C deviates from these assumptions indicating that C contains genuine information about the correlations between the different cryptocurrencies. From here, Louvain community detection method is applied as a clustering mechanism and $15$ community groupings are detected. Finally, PCA is completed on the standardised returns of each of these clusters to create a portfolio of cryptocurrencies for investment. This method selects a portfolio which contains a number of high value coins when compared back against their market ranking in the same year. In the interest of assessing continuity of the initial results, the method is also applied to a smaller dataset of the top $50$ cryptocurrencies across three time periods of $T = 125$ days, which produces similar results. The results obtained in this paper show that these methods could be useful for constructing a portfolio of optimally performing cryptocurrencies.
\end{abstract}

\keywords{Financial Markets \and Graph Theory \and Cryptocurrencies, \and Correlation Matrix \and Community Structures \and Principal Component Analysis}
\vspace{.08in} 
\noindent \textbf{AMS Subject Classification:} 91G45, 62H25, 68R10, 05C85

\section{Introduction}
In recent years, cryptocurrencies have become an increasingly popular investment opportunity \cite{alessandretti2018anticipating}. Since its introduction in 2009, the cryptocurrency investment market has been evolving quite quickly as it has experienced dramatic rises and falls in the space of a few years \cite{makarov2020trading}. Rather than investing in traditional stocks and shares, investors are now opting to invest in the thousands of cryptocurrencies that are available. Coinmarketcap\footnote{Coinmarketcap.com}, shows the daily statistics on over 2,000 cryptocurrencies. With such a wide variety of currencies available for investment, an investor must choose wisely if the aim is to optimise their returns.

Throughout this paper, cryptocurrencies are analysed with the aim of constructing an optimal investment portfolio. Inspired by the recent work of Chaudhari et al.\cite{chaudhari}, we will analyse cryptocurrencies based on their  groupings formed using the Louvain community detection algorithm. Chaudhari et al.\cite{chaudhari} analysed cross-correlation dynamics and community structures within 59 cryptocurrencies over 150 days. In this paper, the methods applied by Chaudhari et al. will be extended in a number of different ways. Firstly, these methods are applied to a larger dataset with over three times the amount of cryptocurrencies (146 cryptocurrencies) over a larger time period (365 days).

Secondly, the community structures are further enhanced by incorporating Principal Component Analysis (PCA) into this study.
PCA is a technique used for dimensionality reduction in data analysis. Using linear combinations of input variables, PCA reduces a large number of variables into a smaller number of factors while still maintaining most of the variability within the data \cite{hargreaves2015selection}. In this paper, this method is applied to the returns of cryptocurrencies based on their community groupings. The cryptocurrencies which account for the highest amount of variability within the first principal component are deemed to be the \textit{Leading Cryptocurrency}. These \textit{leading cryptocurrencies} are combined and analysed to discover if these methods can be applied to construct a portfolio of best performing cryptocurrencies for investment. 

Thirdly, the continuity and robustness of the methods are compared and analysed. \textit{Leading Cryptocurrencies} obtained are compared against market rankings to understand if these methods are useful for detecting stable cryptocurrencies which are likely to stay in a high rank after extended periods of time thus, helping to create an attractive investment portfolio. 

This paper is organised as follows: Section \ref{lit_review} sees a review of the relevant literature surrounding the application of topics in financial and non-financial analysis. The method of obtaining data for this analysis is detailed in section \ref{data}. This section gives details on the pre-processing techniques used to address  incomplete records included in the data. Section \ref{method} details the methods used throughout this analysis. These methods include RMT, community detection and PCA. These methods are then repeated on a smaller dataset.  Finally, section \ref{results} looks at the results obtained and conclusions are discussed in section \ref{conclusion}.

\section{Literature Review}\label{lit_review}
Cryptocurrencies first began trading in 2009 with Bitcoin being one of the earliest coins with many more following on from it. While cryptocurrencies are still quite new on the trading market, there has already been plenty of research emerging into this relatively new area of investment in an attempt to further understand the dramatic rises and falls seen in the market and whether their co-movements could be useful \cite{makarov2020trading}. 
Stosic et al. \cite{stosic2018collective} consider the collective behaviour of cryptocurrency price changes. They analyse the correlations between 119 cryptocurrencies over a period of 200 days and identify hierarchical structures and groupings of cryptocurrency pairs \cite{stosic2018collective}. Through these correlations, they discover the existence of network and community structures in the minimum spanning tree formed from the data. Stosic et al. conclude that these behaviours can be useful for building cryptocurrency investment portfolios. \cite{stosic2018collective}. Liew et al.\cite{liew2019cryptocurrency} carry out a similar analysis, during which they consider the cross-correlation dynamics within cryptocurrencies. They also use PCA and machine learning methods to predict cryptocurrency prices. As part of their analysis, they perform tests across different numbers of cryptocurrencies and across different time periods to avoid bias in their results.

There has also been research into the effect of Bitcoin on other cryptocurrencies. Hu et al. utilise PCA to investigate the relationship between Altcoin \textit{(coins alternative to Bitcoin)}   and Bitcoin returns \cite{hu2019cryptocurrencies}. For the purpose of their analysis, they only choose currencies that have at least 2 years of time series data to ensure that only coins with sufficiently complete data is included. They show that the principal component of Altcoins daily and monthly returns is highly correlated with Bitcoin returns.

Punceva  et al.\cite{punvceva2018bitcoin} consider community groupings within cryptocurrencies. However, unlike other studies, they specifically consider trust communities\footnote{\textit{Trust communities} are created using the ratings of users after interacting with each other. These ratings are used as weights for creating such networks.} within cryptocurrencies. The data used in their study contains user ratings based on interactions with cryptocurrencies and focuses on the Bitcoin OTC and Bitcoin-Alpha networks. They find that four different community detection algorithms (\textit{Louvain, Infomap, Walktrap and Leading Eigenvector}) perform well and give optimal modularity. They conclude that these community structures exist and can help to decide strategies of how to interact on cryptocurrency exchange platforms \cite{punvceva2018bitcoin}. 

Cryptocurrencies are clearly a new class of investible instruments \cite{hu2019cryptocurrencies}, therefore one should also consider research into other financial assets available for investment. There have been similar studies on traditional stocks and shares, using PCA and community detection to build optimal investment portfolios.
Hargreaves et al. \cite{hargreaves2015selection} show how PCA can be used to select optimally performing stocks. They take 22 variables for over 100 stocks. Through their analysis, they reduce the 22 variables to 4 \textit{(Return on Investment, Return on Equity, Book Value per Share and Revenue per Share)}. These 4 variables were used to identify optimal stocks for investment \cite{hargreaves2015selection}. Zahedi et al \cite{zahedi2015application} used similar methods while analysing stock prices in the Tehran stock exchange. They concluded that these stock prices can be predicted using artificial neural networks and PCA. 

Conlon et al. has also contributed to the literature in the area of portfolio analysis. In one paper,  Conlon et al. explore the use of Random Matrix Theory for portfolio optimisation of hedge funds \cite{conlon2007random}. By using hedge fund returns to construct a cross-correlation matrix, it is seen that strategy information can be deduced from the deviating eigenvalues of  Random Matrix Theory predictions. The cross-correlation matrix is cleansed using this strategy information, which improves the difference between the predicted risk and actual risk of the portfolio. Thus, they deduce that the filtered correlation matrix is useful for portfolio optimisation \cite{conlon2007random}. In another study, Conlon et al. also analysed the cross-correlations in financial time-series data from two different stock markets. The eigenvalues of the cross-correlations, formed from random subsets of S\&P equities and members of the Dow Jones Euro Stoxx 50 indices, are analysed with various models. Using sliding windows of time intervals of 200 days, Conlon et al. analyse how the correlation matrices and eigenvalue structures change over time \cite{conlon2009cross}.


Network analysis and community detection has been researched in the area of finance and investment. Networks, such as biological, social, technological and informational, are systems which can be represented in the form of graphs where objects are represented by nodes and their interactions define how the nodes are connected \cite{newman2012communities}. Community detection is a method of clustering networks which can give insights into natural divisions or clusters in our data \cite{lu2015parallel}.

Fenn et al. investigated community dynamics within the exchange-rate market and how they can provide insight into the correlation structures and interactions between exchange-rates \cite{fenn2012dynamical}. Their results indicated that the exchange rates located in the centre of a community structure have strong influence on other exchange rates within the community. This can be used to show which exchange rates dominate a market at a given time \cite{fenn2012dynamical}. Fenn et al. concluded that this type of analysis could also be applied to other asset classes.

Numerous community detection algorithms have been developed. The  \textit{Louvain} method is one of which that is investigated in this paper and is explained in more detail in  Section \ref{CommunityDetect}.
Louvain community detection algorithm has been previously applied in studies investigating prediction of stock prices. 
 
Patil et al. explored this idea by comparing the use of graph based deep learning models (using graph convoluted neural networks), graph-based machine learning models (using Louvain community detection algorithm) and traditional statistical approach (Auto Regressive Integrated Moving Average (ARIMA)), for building a prediction model. They concluded that graph-based models outperformed a traditional statistical approach for building a prediction model \cite{patil2020stock}. Wang et al.\cite{wang2015correlation} used Louvain community detection while investigating the correlation structure and dynamics of international real estate securities markets. They analysed markets from 2006-2012 and found that using Louvain community detection on planar maximally filtered graphs showed that national markets clustered together according to their geographical locations\cite{wang2015correlation}.

The Louvain method has also been applied in the area of health informatics. Surien et al. examined Twitter posts using Louvain community detection while analysing opinions about the human papillomavirus (HPV) vaccines. They found that the use of Louvain community detection, alongside topic modelling, was a successful method to characterise Twitter communities for monitoring online opinions of public health applications \cite{surian2016characterizing}. They concluded that this method could be useful in the future to identify online communities that can be easily influenced by the negative opinions about public health interventions \cite{surian2016characterizing}. 

\section{Data}\label{data}
In this study, cryptocurrency data was obtained using the \textit{crypto} package in R. The \textit{crypto} package retrieves historical opening, high, low and closing values for all cryptocurrencies available on the website: {https://coinmarketcap.com/} \cite{CryptoR}. 

A historical snapshot of the top 200 cryptocurrencies as at 29\textsuperscript{th} December 2019 was extracted from the {https://coinmarketcap.com/}. The historical closing values were then extracted for these top 200 coins using the \textit{crypto} package \cite{CryptoR}. A full years' worth of data was considered for this analysis in order to allow for seasonality. 

Data was extracted for the top 200 coins from 1\textsuperscript{st} January 2019 to 31\textsuperscript{st} December 2019. In this time period, not all cryptocurrencies had been trading consistently. Coins which did not have a full years' worth of data were excluded from the dataset. This issue occurred as many of the coins in the top 200 may have only started trading in the middle of 2019. (e.g. LEO is a coin which featured 14\textsuperscript{th} in our top 200 but only began trading on the 20th of May). Of the 200 coins, 146 had price records for everyday of the year and the remaining coins were removed from the data. 

\section{Method}\label{method}
\subsection{Calculating the Cross-Correlations}\label{cross-corr}

Some pre-processing must be completed on the closing prices in order to derive a correlation matrix $C$. Firstly, the returns of the cryptocurrencies are calculated. Let $P_{i}(t)$ be the closing price of cryptocurrency $i$ at a given time $t$, with $i=1,2,...,N$. Daily log returns are calculated using:
\begin{equation}\label{log_returns}
G_{i}(t) = ln(P_{i}(t + \Delta t))-ln(P_{i}(t))
\end{equation}
where $\Delta t$ is a specified time interval. As the data contains daily closing prices, $\Delta t = $ 1 day.
In a similar fashion to other papers analysing cross-correlations of price returns, normalised returns are used for constructing the cross-correlation matrix $C$. As the levels of volatility (i.e. standard deviations) differ between cryptocurrencies, a normalised return allows for the standardisation of  volatility and ensures that results are independent of varying levels of volatility  \cite{conlon2009cross}\cite{shen2009cross}. The normalised return is given by:
\begin{equation}\label{normalised_return}
    g_{i}(t) \equiv \frac{G_{i}(t)-\langle G_{i}\rangle}{\sigma_{i}} 
\end{equation}
Where $\langle G_{i}\rangle$ refers to the average $G_{i}$ over a time window $T$. $\sigma_{i}$ is the standard deviation of $G_{i}$ such that $\sigma_{i}\equiv\sqrt{\langle G_{i}^{2}\rangle - \langle G_{i} \rangle^{2}}$. From here, the cross-correlation matrix $C$ can be computed whose elements are of the following form:
\begin{equation}\label{Correlation_Matrix}
    C_{ij} = \langle g_{i}(t)g_{j}(t) \rangle
\end{equation}
$C_{ij}$ measures the correlation between the normalised returns of two cryptocurrencies $i$ and $j$. A value of $C_{ij} = 1$ indicates a perfectly positive relationship between the normalised returns of cryptocurrencies $i$ and $j$, and a value of $C_{ij} = -1$ defines a perfectly negative relationship between the normalised returns of $i$ and $j$. Finally, a value of $C_{ij} = 0$ means that there is no correlation between the normalised returns of coins $i$ and $j$ 
\cite{stosic2018collective}. $C$ is a $N \times N$ matrix where $N=$ no. of cryptocurrencies. In our case, $N = 146$. 
\par
\begin{figure}[!ht]
    \centering
    \includegraphics[width=0.8\textwidth]{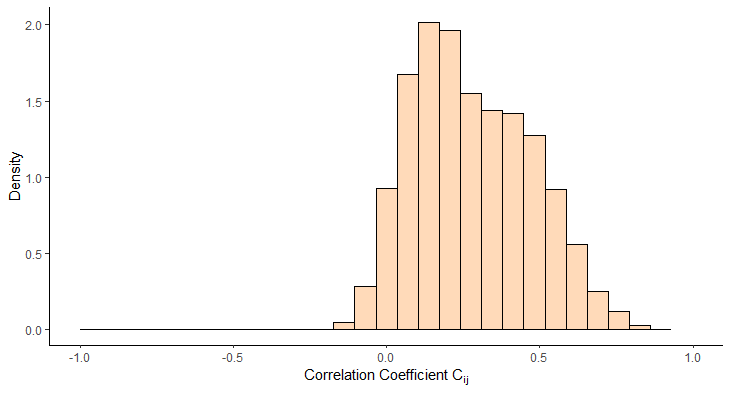}
    \caption{Distribution of Correlation Coefficients $C_{ij}$.}
    \label{fig:correlations}
    \centering
\end{figure}

\subsection{Random Matrix Theory Statistics}\label{RMT}
The unpredictability of financial returns over time has been commented on many times (
see, e.g., \cite{mantegna1999hierarchical}). 
Random Matrix Theory (RMT) is a common technique used to understand the changes in correlations of investments. RMT tests the statistics of the correlation matrix $C$ against the properties of a random matrix $R$. If the properties of $C$ deviate from those of our random matrix $R$, then this indicates that the correlation matrix $C$ contains information about genuine correlations \cite{plerou2002random}. The eigenvalues and eigenvectors calculated from the correlation matrix provide information about the collective behaviour of the system of cryptocurrencies \cite{stosic2018collective}.

The use of Random Matrix Theory for analysing interacting systems originated from the work of Wigner \cite{wigner1993class} who applied it to systems of interaction while studying the energy levels of nuclei. He assumed that the components of the nucleus can be modelled as random \cite{stosic2018collective}\cite{plerou2000random} such that they follow a Gaussian distribution with mean zero and unit variance. The results achieved by Wigner were later extended to financial markets by Laloux et al. \cite{laloux1999noise} and Plerou et al. \cite{plerou1999universal} who used it to analyse the cross-correlations of stock market returns. Both found that the method was useful for discovering 
information from the cross-correlation matrices and could be used for portfolio optimisation. Since then, RMT has become a popular tool for investigating cross-correlations of financial markets\cite{conlon2009cross} and has also been extended to the cryptocurrency markets 
\cite{chaudhari}\cite{stosic2018collective}.

 The calculated cross-correlation matrix $C$ can be represented in the following matrix notation:
\begin{equation}
    C = \frac{1}{T}GG^{T}
\end{equation}
where $G$ is a matrix of size $N \times T$. As discussed by Hu et al. \cite{hu2019cryptocurrencies}, eigenvalues, $\lambda_{i}$, and eigenvectors, $\hat{v_{i}}$, are obtained from the matrix $C$ using the following:
\begin{equation}
    C\hat{v_{i}} = \lambda_{i}\hat{v_{i}} 
\end{equation}
In order to understand if $C$ confines to the assumptions of RMT, it is compared to the matrix $R$ which is a random cross-correlation matrix of the following form:
\begin{equation}
    R = \frac{1}{T}AA^{T}
\end{equation}
In which $A$ is a $N \times T$ matrix with zero mean and unit variance \cite{plerou2002random}. $R$ belongs to a class of random matrices in multivariate statistics called Wishart matrices \cite{plerou2002random}. One property of these matrices is that as $N \xrightarrow{} \infty$, $T \xrightarrow{} \infty$, with $Q\equiv T/N \geq 1$, the distribution of eigenvalues ($\lambda_{i}$) of $R$ can be represented in the following form:
\begin{equation}
    P_{R}(\lambda) = \frac{Q}{2\pi\sigma^{2}}\frac{\sqrt{(\lambda_{+}-\lambda)(\lambda - \lambda_{-})}}{\lambda}
\end{equation}
$\lambda_{+}$ and $\lambda_{-}$ are the upper and lower bounds of the eigenvalues, $\lambda_{i}$, of $R$ such that $\lambda_{-} \leq \lambda_{i} \leq \lambda_{+}$ for $i = 1,2,...,N$. $\sigma^{2} = 1$ due to the normalisation completed on the price returns while accounting for volatility \cite{plerou2002random}\cite{laloux2000random}. This give the following formula for calculating expected eigenvalue bounds of a Wishart matrix:
\begin{equation}
    \lambda_{+},\lambda_{-} = 1+\frac{1}{Q} \pm 2\sqrt{\frac{1}{Q}}
\end{equation}
This assumed distribution  is compared against the calculated eigenvalues from the correlation matrix $C$. The extracted data consists of $N=146$ coins over a period of $T=365$ days, such that the Quality Factor\cite{bouchaud2000theory}, $Q=T/N = 365/146 = 2.5$. This results in eigenvalue bounds of $\lambda_{-}=0.135$ and $\lambda_{+}=2.665$. 
\begin{figure}[!ht]
    \centering
    \includegraphics[width=1\textwidth]{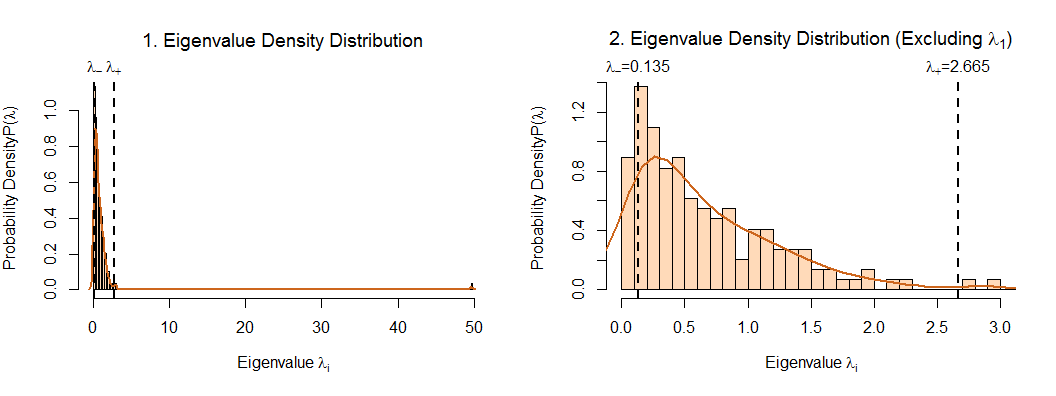}
    \caption{Eigenvalue Distribution.}
    \label{fig:Eigenvalue_Dist}
    \centering
\end{figure}
Figure \ref{fig:Eigenvalue_Dist} shows the distribution of calculated eigenvalues from the correlation matrix $C$. Figure \ref{fig:Eigenvalue_Dist}$(1)$ shows the distribution of all eigenvalues and the second graph, Figure \ref{fig:Eigenvalue_Dist}$(2)$ shows the distribution of all eigenvalues excluding $\lambda_{1}=49.66$. From figure \ref{fig:Eigenvalue_Dist} we can see that many of the eigenvalues fall within the range of $\lambda_{-}$ and $\lambda_{+}$. Stosic et al. refer to these eigenvalues of $C$ as ``bulk" eigenvalues such that for RMT predictions,  $\lambda_{i} \in \lambda_{bulk}$ fall within the range $\lambda_{-}<\lambda_{bulk}<\lambda_{+}$\cite{stosic2018collective}. While this is true for the ``bulk" of the eigenvalues, we can also see there are several  eigenvalues which fall outside of this range. These deviations from the predicted $\lambda_{bulk}$ suggests that $C$ contains genuine information about the correlations between the different cryptocurrencies \cite{stosic2018collective}\cite{plerou2000random}.

While it has been shown that the eigenvalues of the correlation matrix $C$ deviate from the RMT predictions, the distribution of eigenvector components must also be considered and analysed. Plerou et al.\cite{plerou2002random} explain that, under RMT, the eigenvector components ($u_{l}^{k}, l=1,...,N, k=1,...,N$) of a correlation matrix $C$ are expected to follow a Gaussian distribution with zero mean and unit variance such that \cite{plerou2002random}:
\begin{equation}
    P_{RMT(u)}=\frac{1}{\sqrt{2\pi}}\exp{-\frac{u^{2}}{2}}
\end{equation}

\begin{figure}[!ht]
    \centering
    \includegraphics[width=1\textwidth]{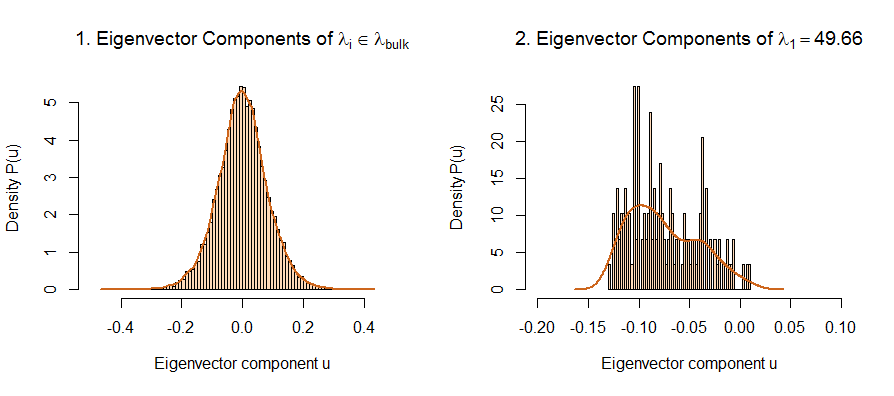}
    \caption{Eigenvector Distribution.}
    \label{fig:eigenvector distribution}
    \centering
\end{figure}
Figure \ref{fig:eigenvector distribution} shows the distribution of eigenvector components corresponding to the eigenvalues, $\lambda_{i}$, of $C$. Figure \ref{fig:eigenvector distribution}$(1)$ shows the distribution of the eigenvector components corresponding to the bulk eigenvalues $\lambda_{bulk}$ such that $\lambda_{-}<\lambda_{bulk}<\lambda_{+}$. The second graph, Figure \ref{fig:eigenvector distribution}$(2)$, shows the distribution of eigenvector components for the largest eigenvalue $\lambda_{1} \notin \lambda_{bulk}$. For the graph showing the eigenvector components corresponding to the bulk eigenvalues, $\lambda_{bulk}$, we can see that these appear to conform to RMT predictions of a Gaussian distribution. However, for the largest eigenvalue $\lambda_{1}$, this is not the case. In figure \ref{fig:eigenvector distribution}$(2)$ we can see that its eigenvector components do not follow a similar distribution. This deviation from the predicted model suggests that the cryptocurrencies ``move" together and that there exists correlations which influence the entire system \cite{plerou2000random}.

A Kolmogorov-Smirnov (KS) test was also used to compare the eigenvector components from $\lambda_{i} \in \lambda_{bulk}$ and $\lambda_{i} \notin \lambda_{bulk}$. The KS test rejected the null hypothesis when comparing the eigenvector components from inside and outside of the eigenvalue bulk. This provides further evidence that deviating eigenvalues do not conform to the predictions of RMT.    

\begin{figure*}[h!]
\begin{center}
\begin{tikzpicture}
    [line cap=round,line join=round,x=2cm,y=2cm,
     spy using outlines={rectangle,lens={scale=1.5}, size=4cm, connect spies},
     decoration={brace,amplitude=2pt}]
\node {\includegraphics[width=0.9\textwidth]{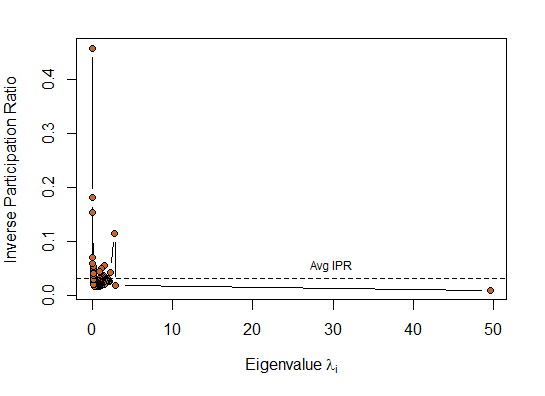}};
   \spy [blue] on (-2,-0.93)
             in node [left] at (1.5,0.8);
\end{tikzpicture}
    \caption{Inverse Participation Ratio (IPR) of eigenvector components of C with (inset) eigenvalues around the average IPR= 0.033.}
    \label{fig:IPR}
\end{center}
\end{figure*}

While analysing the components of the eigenvectors, it is also important to consider the inverse participation ratio (IPR). The IPR is used for quantifying the number of components that participate significantly in each eigenvector \cite{conlon2007random}. It works by computing the inverse of the number of eigenvector components that contribute to each eigenvector component using the following formula:
\begin{equation}
    I^{k} \equiv \sum_{l=1}^{N}[u_{l}^{k}]^{4}
\end{equation}
where $u_{l}^{k}$, $l=1,...,N$ are the components of eigenvector $\boldsymbol{u}^{k}$ \cite{conlon2007random}\cite{plerou2002random}. Figure \ref{fig:IPR} shows the IPR of the eigenvector components of our correlation matrix $C$. The graph shows the average IPR $=0.033$ in which most of the eigenvalues appear to fluctuate around. As mentioned by Stosic et al., these fluctuations show that almost all of the cryptocurrencies contribute to these eigenvectors \cite{stosic2018collective}. This further shows that $C$ does not conform to the RMT predictions and hence, it contains genuine information on the correlations of cryptocurrencies. 

\subsection{Financial Markets as Network Structures}\label{networkstructures}
In this paper, the interactions within financial markets are modelled using network procedures \cite{mantegna1999hierarchical}. The cryptocurrencies in this study can be represented by a network structure such that the nodes represent each cryptocurrency and the edges connecting them represent the correlations between their standardised returns \cite{mantegna1999hierarchical}\cite{namaki2011network}. 

Mantegna \cite{mantegna1999hierarchical} explains that, when forming a network using cross-correlations, it is not correct just to use correlation coefficients from the correlation matrix $C$ to form the network as it does not fulfil the three axioms that define a metric for building networks \cite{mantegna1999hierarchical}. He further explains the method of using a distance metric, which has been applied in other studies while modelling financial returns with network structures\cite{chaudhari}\cite{stosic2018collective}\cite{wang2015correlation}. The distance metric is a transformation of the correlation matrix $C$ into a distance matrix $D$ using the following:
\begin{equation}\label{Distance}
D_{ij}=\sqrt{2(1-C_{ij})}
\end{equation}
Where $C_{ij}$ and $D_{ij}$ are the elements of the matrices $C$ and $D$. The transformation of $C$ into the distance matrix $D$ will now fulfil the three axioms of a distance metric: 1. $D_{ij} = 0 \iff i=j$, 2. $D_{ij}=D_{ji}$, 3. $D_{ij} \leq D_{ik}+D_{kj}$ \cite{mantegna1999hierarchical}. This distance metric can now be used to build a network model of the cryptocurrency returns.

Following from the work of Mantegna \cite{mantegna1999hierarchical}, a minimum spanning tree (MST) approach was used to form the correlation-based network of the 146 cryptocurrencies in this analysis. The idea of using MST for modelling financial markets was first introduced by \cite{mantegna1999hierarchical} while analysing the hierarchical structure of the US stock market and has become a commonly used network structure for analysing the correlations in financial time series\cite{chaudhari}. Using an $N \times N$ distance matrix $D$, a MST connects $N$ nodes using $N-1$ edges in such a way that the sum of all distances is the minimum  \cite{mantegna1999hierarchical}\cite{wang2015correlation}. $D_{ij}$ represents the weight on the edge connecting cryptocurrencies (nodes) $i$ and $j$.

Minimum spanning trees can be implemented using different algorithms.  \textit{Prim} and  \textit{Kruskal} are commonly used algorithms for modelling hierarchical structures of financial returns. Prim's algorithm forms a minimal spanning tree by ``growing" a network from a single root node. Whereas the Kruskal algorithm forms a minimum spanning tree by considering the weight of every edge and ``growing" in clusters \cite{huang2009comparison}. Huang et al. compared the two algorithms while studying the Shanghai and Shenzhen 300 index stock market exchange \cite{huang2009comparison}. 
They concluded that Kruskal was more suitable for sparse-edged network structures, whereas Prim's algorithm was more suitable for minimum spanning trees with dense edges \cite{huang2009comparison}. As the network formed from the distance matrix $D$ had a $N=146$ nodes and $\sim10,000$ vertices, Prim's algorithm was used to create the minimum spanning tree of $146$ cryptocurrencies. 
\subsection{Community Detection}\label{CommunityDetect}
When the MST is constructed using Prim's algorithm, the collective behaviour between cryptocurrencies can be analysed using community structures. A community structure exists in a network when the vertices can be divided into groups in such a way that the groups of vertices are compactly connected internally but the different groups are sparsely connected \cite{wang2015correlation}. Similar to minimum spanning trees, there are many different community detection algorithms. Following on from the methods applied by Chaudhari et al. \cite{chaudhari}, the \textit{Louvain} community detection method is used for investigating the community structures within the MST of $N=146$ cryptocurrencies. 

The \textit{Louvain} method, developed by Blondel et al., is a heuristic method which creates communities based on modularity optimisation \cite{blondel2008fast}. Newman\cite{newman2016community} describes modularity as a benefit function, that is a measure of the quality of divisions of a network into communities. Every partition has a modularity value which measures the density of links inside communities compared to links between communities \cite{blondel2008fast}. The Louvain algorithm creates communities by initialising nodes to a community of size one. Using a ``bottom-up" approach, the clusters are sequentially combined with other clusters that produce the greatest gain in modularity (if a positive gain exists). This process is continued until there are no further gains and modularity has been optimised \cite{surian2016characterizing}. 

\section{Results}\label{results}
\begin{figure}[!ht]
    \centering
    \includegraphics[width=0.9\textwidth]{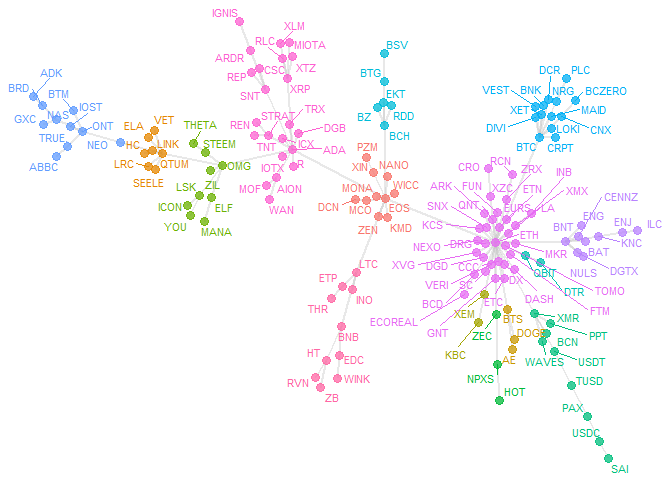}
    \caption{Minimum Spanning Tree: Community detection - 15 Communities.}
    \label{fig:MST-Louvain-Fulldataset}
    \centering
\end{figure}

The Louvain community detection algorithm was applied to the minimum spanning tree. 15 distinct communities were formed from this as shown in figure \ref{fig:MST-Louvain-Fulldataset}. The different colours represent the 15 different communities formed. Of the 15 communities, 4 had 3 or less members belonging to them. The cryptocurrencies which made up these small communities $(<4\text{ members})$ were generally all low valued coins and can all be discounted from the next part of the analysis. The remaining communities had sizes ranging from $6-31$.

Once the Louvain community detection method was applied, PCA was performed onto the community groupings to identify the cryptocurrency which explained the largest amount of variance within the dataset.

Using the normalised returns data calculated using equation (\ref{normalised_return}), the dataset of $146$ coins was subset into $15$ groups based on their community groupings detected with the Louvain algorithm. From here, PCA was performed on each of the community groupings to determine which cryptocurrency explained the highest amount of variance within each of the groups of returns. The cryptocurrency which explained the highest amount of variance within the first principal component of the returns data is detailed in Table \ref{Table:Louvain and PCA for 146} as `Leading Coin (PCA)'. 

Table \ref{Table:Louvain and PCA for 146} shows details of the sizes of communities and their leading cryptocurrency from the group. 
The `leading coins' were referenced back against the historical snapshot of top 200 coins as at 29\textsuperscript{th} of December 2019. This ranking is a historical snapshot of the coins ranking by market capitalisation and is shown in the fourth column of table \ref{Table:Louvain and PCA for 146}. We can see that 7 of the leading coins \textit{(EOS, Tether, Bitcoin Cash, Bitcoin, Ethereum, Cardano, Litecoin)} lie in the top 20. \textit{Qtum, OmiseGO, Ontology} lie between the 20-60 ranking and one coin, \textit{Bancor}, was the lowest ranked leading coin from the analysis. As 7 of the 11 coins lie in the top 20 leading coins by market cap, this shows that the methods applied in this paper could be useful for creating cryptocurrency investment portfolios.

The table also shows the ranking of those coins 6 months later as at 28\textsuperscript{th} June 2020. Comparing these results with their ranking 6 months later, we see that the 7 coins in the top 20, are remaining in the top 20 ranking. While the other coins, still stay close to the original ranking from the 29\textsuperscript{th} December 2019. The lack of deviations from their ranking after 6 months further shows that the methods applied in this paper could be useful for creating cryptocurrency investment portfolios.

\begin{table}[h!]\caption{Louvain Community Detection $\And$ PCA of 146 Cryptocurrencies}
\centering
\resizebox{\textwidth}{!}{
 \begin{tabular}{||c|c|c|c|c||} 
 \hline
 \textbf{Community}& \textbf{Community Size} &\textbf{Leading Coin (PCA)} &\textbf{Ranking (29/12/2019)} & \textbf{Ranking (28/06/2020)}\\ [0.4ex] 
 \hline\hline
    1 & 10 & EOS & 7 & 9 \\ 
    2 & 7 & Qtum & 41 & 62 \\ 
    5 & 9 & OmiseGO & 58 & 48 \\ 
    7 & 9 & Tether & 4 & 3 \\ 
    9 & 6 & Bitcoin Cash & 5 & 5 \\ 
    10 & 13 & Bitcoin & 1 & 1 \\ 
    11 & 10 & Ontology & 29 & 32 \\ 
    12 & 9 & Bancor & 184 & 105 \\ 
    13 & 31 & Ethereum & 2 & 2 \\ 
    14 & 22 & Cardano & 13 & 11 \\ 
    15 & 10 & Litecoin & 6 & 7 \\ [1ex]
 \hline
 \end{tabular}}
 \label{Table:Louvain and PCA for 146}
\end{table}

In order to test if there is continuity in the results obtained, the data was subdivided into a smaller dataset of the top 50 cryptocurrencies. Over three time periods of $T=125$ days, the methods were reapplied to investigate if similar results could be obtained with a smaller dataset. With this time window, the $Q$ factor\footnote{Q factor is the ratio of the number of observations (days) to the number
of assets \cite{bouchaud2000theory}. It is used to calculate eigenvalues bounds which are compared against eigenvalues of correlation matrix $C$ to test RMT assumptions.} is still maintained such that $Q = 125/50 = 2.5$. 
The three subsets of $T=125$ days cover the following date ranges:
\begin{enumerate}
\centering
    \item $01/01/2019-05/05/2019$
    \item $01/05/2019-02/09/2019$
    \item $29/08/2019-31/12/2019$
\end{enumerate}
The data was pre-processed using the same methods to obtain the normalised return as detailed in equations (\ref{log_returns}) and (\ref{normalised_return}) and cross-correlation matrices $C_{1}$, $C_{2}$, $C_{3}$ were computed using the three sets of standardised returns.
\par 
From this, the eigenvalues and eigenvectors were calculated from each correlation matrix and analysed using RMT statistics in a similar fashion as above. The distribution of the eigenvalues was that of what was seen for the larger dataset of $T=365$ days. For all time periods, the majority of eigenvalues, $\lambda_{i}$ fell within the bulk with the same eigenvalue bounds of $\lambda_{-}=0.135$ and $\lambda_{+}=2.665$ such that $\lambda_{-} \leq \lambda_{bulk} \leq \lambda_{+}$. There were also deviations from this, with some eigenvalues falling outside of the bulk $(\lambda_{i} \notin \lambda_{bulk})$. The eigenvector components for these eigenvalues were also analysed. Similar to our large dataset, the eigenvector components for eigenvalues within the bulk,  $\lambda_{bulk}$, all appeared to conform to a Gaussian distribution. Whereas, for the eigenvalues outside of the bulk, $(\lambda_{i}\notin\lambda_{bulk})$, the distribution of their eigenvector components also deviated from a Gaussian normal distribution. A Kolmogorov-Smirnov (KS) test confirmed this deviation as the KS test rejected the null hypothesis when comparing the eigenvector components from inside and outside of the eigenvalue bulk for each of the three correlation matrices. Similar to the cross-correlation matrix $C$ for $N=146$ cryptocurrencies, this confirms that cross-correlation matrices $C_{1}$, $C_{2}$ and $C_{3}$ do not conform to RMT predictions and contain genuine information on cross-correlations of the top 50 cryptocurrencies. 
 
 \begin{table}[h!]\caption{Top 50 Cryptocurrencies - Louvain Community Detection - No. of communities and their sizes}
\centering

\begin{tabular}{||cc|cc|cc||}
    \hline
    \multicolumn{2}{||c|}{$\boldsymbol{T_{1}}$} & \multicolumn{2}{c|}{$\boldsymbol{T_{2}}$} & \multicolumn{2}{c||}{$\boldsymbol{T_{3}}$} \\
    \textbf{Community} & \textbf{Size} & \textbf{Community} & \textbf{Size} & \textbf{Community} & \textbf{Size}\\[0.5ex]
    \hline
    \hline
    1 & 3 & 1 & 8 & 1 & 4 \\
    2 & 5 & 2 & 5 & 2 & 8 \\
    3 & 8 & 3 & 5 & 3 & 4 \\
    4 & 8 & 4 & 8 & 4 & 3 \\
    5 & 7 & 5 & 11 & 5 & 6 \\
    6 & 8 & 6 & 8 & 6 & 7 \\
    7 & 6 & 7 & 5 & 7 & 4 \\
    8 & 5 & - & - & 8 & 7 \\
    - & - & - & - & 9 & 7 \\

    \hline
\end{tabular}
\label{table:Top50Communities}
\end{table} 
\par
As with the above methods used on $N=146$ cryptocurrencies, three distance matrices were calculated using the cross-correlation matrices $C_{1},C_{2},C_{3}$ and formula (\ref{Distance}). Louvain community detection was applied to the minimum spanning trees formed using the distance matrices and Prim's algorithm.
\par
\begin{figure}[!ht]
    \centering
    \includegraphics[width=0.8\textwidth]{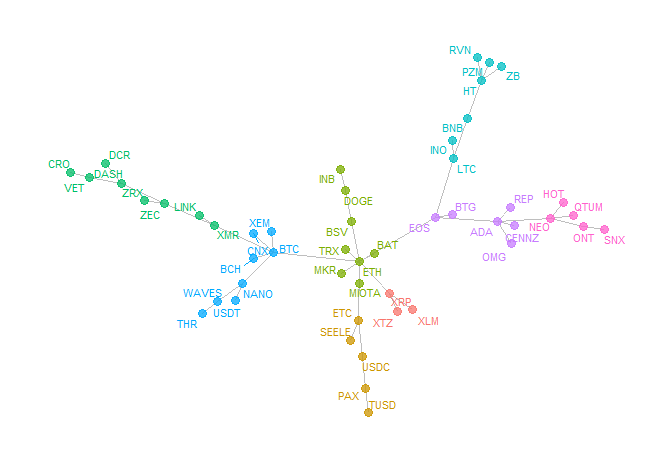}
    \caption{Minimum Spanning Tree with Louvain Communities - Top 50 - Time window $T_{1}$.}
    \label{fig:Top50 Time window 1}
    \centering
\end{figure}
\begin{figure}[!ht]
    \centering
    \includegraphics[width=0.8\textwidth]{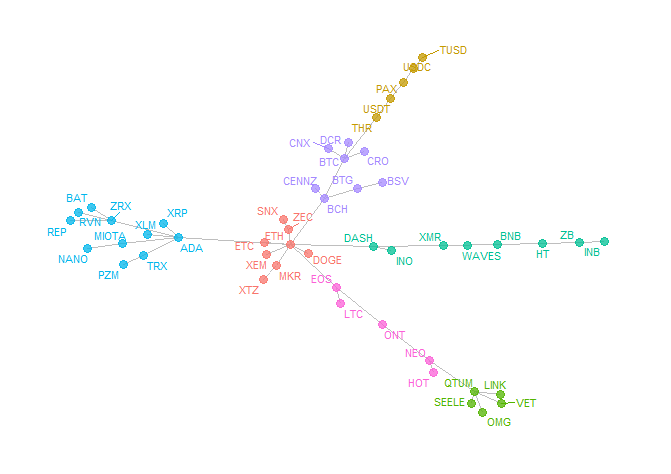}
    \caption{Minimum Spanning Tree with Louvain Communities - Top 50 - Time Window $T_{2}$.}
    \label{fig:Top50 time window 2}
    \centering
\end{figure}
\begin{figure}[!ht]
    \centering
    \includegraphics[width=0.8\textwidth]{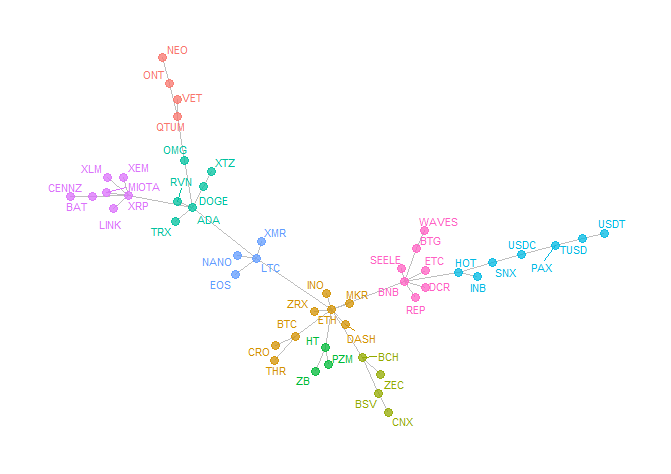}
    \caption{Minimum Spanning Tree with Louvain Communities - Top 50 - Time Window $T_{3}$.}
    \label{fig:Top50 time window 3}
    \centering
\end{figure}

Table \ref{table:Top50Communities} shows the initial results when Louvain community detection algorithm was applied to $T_{1}$, $T_{2}$ and $T_{3}$. The algorithm detected similar numbers of communities across the time window with 8 communities being detected in $T_{1}$, 7 in $T_{2}$ and 9 in $T_{3}$. Table \ref{table:Top50Communities} also details the number of cryptocurrencies in each community with the community sizes appearing to be similar across the three-time windows. Figures \ref{fig:Top50 Time window 1}, \ref{fig:Top50 time window 2} and \ref{fig:Top50 time window 3} show the graphical representation of the Louvain community detection algorithm applied to the minimum spanning trees across the three time windows $T_{1}$, $T_{2}$ and $T_{3}$. Similar to the minimum spanning tree in figure \ref{fig:MST-Louvain-Fulldataset}, the different colours represent the different communities detected by the algorithm.
\par
Following the same methods applied to the larger dataset of 146 cryptocurrencies, PCA was applied to the standardised returns for each of the communities detected to determine the `Leading Coins' for each community.

Table \ref{tab:interfacesMellemEnheder} shows the results of the `Leading Coins' obtained after applying PCA to the standardised returns of the community groupings detected across the three time windows $T_{1}$,$T_{2}$ and $T_{3}$. The table shows all leading coins and their ranking by market capitalisation on 29\textsuperscript{th} December 2019 and 28\textsuperscript{th} June 2020. From the results table we can see that the coins selected as `leading coins' (i.e. the coins which accounted for most of the variability within the first principal component) appear to be stable in their ranking across the 6 months from December 2019 to June 2020. This shows that the method could be used in future for detecting cryptocurrencies which are deemed to be stable and are less likely to drop in ranking. 

\begin{table}[h!]\centering \caption{Results of PCA performed on Louvain communities for Top 50 Cryptocurrencies over three time periods}
\begin{threeparttable}
\centering
\resizebox{0.8\textwidth}{!}{\begin{tabular}{lP{1.4cm}P{1.4cm}ccc}
\toprule
\multirow{2}{*}{\textbf{Leading Coin (PCA)}} &\multirow{2}{*}{\parbox{1.4cm}{\centering \textbf{Rank** 29/12/2019}}} &\multirow{2}{*}{\parbox{1.4cm}{\centering \textbf{Rank** 28/06/2020}}} & \multicolumn{3}{c}{\textbf{Time Window}} \\
\cmidrule{4-6}
            & & &$\boldsymbol{T_{1}}$                &  $\boldsymbol{T_{2}}$             & $\boldsymbol{T_{3}}$ \\
\midrule
Bitcoin         & 1  & 1  & \checkmark       & \checkmark       & \cellcolor{Gray} \\
Ethereum        & 2  & 2  & \checkmark       & \checkmark       & \checkmark \\
XRP             & 3  & 4  & \checkmark       & \cellcolor{Gray} & \checkmark \\
Bitcoin Cash    & 5  & 5  & \cellcolor{Gray} & \cellcolor{Gray} & \checkmark \\
Litecoin        & 6  & 7  & \cellcolor{Gray} & \cellcolor{Gray} & \checkmark \\
Binance Coin    & 8  & 8  & \cellcolor{Gray} & \cellcolor{Gray} & \checkmark \\
Cardano         & 13 & 11 & \checkmark       & \checkmark       & \checkmark \\
Monero          & 16 & 17 & \cellcolor{Gray} & \checkmark       & \cellcolor{Gray} \\
Huobi Token     & 17 & 20 & \checkmark       & \cellcolor{Gray} & \checkmark \\
Neo             & 19 & 21 & \checkmark       & \cellcolor{Gray} & \cellcolor{Gray} \\
Dash            & 28 & 23 & \checkmark       & \cellcolor{Gray} & \cellcolor{Gray} \\
Ontology        & 29 & 32 & \cellcolor{Gray} & \checkmark       & \checkmark \\
Paxos Standard  & 36 & 43 & \checkmark       & \checkmark       & \checkmark \\
Qtum            & 41 & 62 & \cellcolor{Gray} & \checkmark       & \cellcolor{Gray} \\
\bottomrule
\end{tabular}}

\begin{tablenotes}\footnotesize
\item[*]$T_{1} = $Time period of $T=125$ days  from 01/01/2019-05/05/2019
\item[*]$T_{2} = $Time period of $T=125$ days from 01/05/2019-02/09/2019
\item[*]$T_{3} = $Time period of $T=125$ days from 29/08/2019-31/12/2019
\item[**] Ranking refers to ranking by market capitalisation
\end{tablenotes}
\end{threeparttable}
\label{tab:interfacesMellemEnheder}
\end{table}

After PCA was applied to the community groupings, there were three `Leading Coins' which were common across the three windows. \textit{Ethereum}, \textit{Cardano} and \textit{Paxos Standard} were all common leading coins detected when PCA was applied to the returns of the cryptocurrencies detected with Louvain community detection. Of the 14 coins detected as leading coins with PCA, 7 of them appeared as leading coins in other time windows. Whereas the remaining 7 only appeared as a leading coin of communities in one of the time windows. This could indicate that there may not be continuity in the methods applied in this paper. Another interesting finding was noticed while analysing the leading coins discovered from the community structures formed in $T_{3}$. 7 of the 9 leading coins detected in this time window are within the Top 20 ranking as at 29\textsuperscript{th} December 2019 and as at 28\textsuperscript{th} June 2020. This could indicate that the method of using community detection and PCA could be useful for creating stock portfolios based on the data from the later part of the year.  

\section{Conclusion}\label{conclusion}
This paper investigates the collective behaviour of 146 cryptocurrencies from 1\textsuperscript{st} January 2019 to 31\textsuperscript{st} December 2019. The data was pre-processed, and standardised returns were obtained in order to calculate the correlation matrix. This correlation matrix was analysed and compared against the assumptions of Random Matrix Theory. The deviations from RMT predictions are in line with other research \cite{chaudhari}\cite{stosic2018collective}\cite{plerou2002random} which shows that the correlation matrix contains genuine information between the interactions of cryptocurrencies. From here, the minimum spanning trees formed from the cross-correlations were analysed using \textit{Louvain} community detection. Principal Component Analysis was then used to find `Leading Coins' with the aim of constructing an investment portfolio of well performing cryptocurrencies. 

The results of this analysis selected several high-valued coins which continued to remain high value when their ranking was compared six months later to 28\textsuperscript{th} June 2020. This shows that the methods applied in this paper could be useful for constructing a portfolio of consistently optimal cryptocurrencies for investment.

The above methods were replicated on a smaller dataset of the top 50 cryptocurrencies of three smaller time windows of $T=125$ days to understand if there is continuity in the results obtained on the larger dataset. The replication of the methods on the smaller datasets showed some continuity, in that several higher valued cryptocurrencies (within the top 20 ranking) were chosen and found across the results of the other time windows. However, there was also cryptocurrencies which appeared only once as `Leading Coins' across the three-time windows. This shows that further testing would be required to validate the results. Also, of the 9 leading coins selected in the third window $T_{3}$, 7 of them were in the top 20 ranking for December 2019 and June 2020. This result could indicate that data from the later part of the year could be more successful for selecting high-valued cryptocurrencies using the methods of this paper.

Finally, while the methods applied in this paper did produce some optimal results, there were also some deviations as the methods were applied over smaller time windows. For future work, these methods would need to be repeated on different time periods and time window lengths to understand if this could be a reliable method for creating investment portfolios.

\bibliographystyle{elsarticle-num}
\bibliography{./Bib2}
\end{document}